\documentclass[twocolumn]{aastex631}

\begin{document}

\title{C14 Automatic Imaging Telescope Photometry of GJ~1214}

\correspondingauthor{G.W.H.}
\email{gregory.w.henry@gmail.com}

\author[0000-0003-4155-8513]{Gregory W.\ Henry}
\affil{Center of Excellence in Information Systems, Tennessee State University, Nashville, TN 37209 USA}

\author[0000-0003-4733-6532]{Jacob L.\ Bean}
\affil{Department of Astronomy \& Astrophysics, University of Chicago, Chicago, IL 60637, USA}



\begin{abstract}
GJ~1214b is the highest signal-to-noise sub-Neptune for atmospheric studies. Although most previous transmission spectroscopy measurements have revealed a frustratingly featureless spectrum, \textit{JWST} observations are expected to give new insights to this benchmark planet. We have performed photometric monitoring of GJ~1214 (the host star) to provide context for these observations. We find that GJ~1214 entered a period of relatively high brightness during 2021 and 2022. This implies that the \textit{JWST} MIRI/LRS phase curve observation of GJ~1214b in July 2022 was obtained during an epoch of low activity for the spot-dominated host star. Like previous works, we are unable to definitively identify the star's rotation period. Nevertheless, we confirm that it is likely $\gtrsim$50\,days.

\end{abstract}



\section{Introduction} \label{sec:intro}
The transiting exoplanet GJ~1214b was discovered by \citet{cbi+2009} with the MEarth Project array of eight 0.40~m automated telescopes designed to 
monitor a large number of nearby M dwarfs for transiting exoplanets.  They found GJ~1214b to have a planetary mass of 6.55\,M$_{\oplus}$, a radius of 
2.68\,R$_{\oplus}$, and an orbital period of 1.58\,days. Originally classified as a super-Earth, consideration of GJ~1214b in light of the \textit{Kepler} planet demographics \citep{fulton17,vaneylen18} suggests it is better thought of as a sub-Neptune \citep{bean21a}. Its low density of 1.9\,g/cc implies the presence of a substantial atmosphere \citep{rogers10}.  There has been extensive effort to detect this atmosphere using transmission spectroscopy \citep[e.g.,][]{bean10, croll11, bean11, desert11, berta12, fraine13, kreidberg14, kasper20, orell22, spake22}, with the consensus being that the planet has a featureless spectrum due to high-altitude aerosols.

The 2009 MEarth photometry found the  star to vary in brightness by 2\% on a timescale of several weeks (with a dominant period of 83 days).  \citet{cbi+2009} concluded that starspots  carried around the star by its rotation was the most likely explanation.  \citet{cwh+2011} and \citet{kreidberg14} observed 31 transits between 2009 and 2013 and found that four transits exhibited brightness anomalies as the planet occulted a starspot, confirming the presence of dark spots as the cause of the
star's brightness variability.

Subsequent studies have confirmed the low-amplitude stellar variability but have not been very successful at pinning down the true stellar rotation 
period.  \citet{bcb+2011} analyzed new 2010 MEarth photometry with better  sampling and cadence than the 2009 MEarth discovery observations and found a best period of 53 days.  However, they cautioned that if GJ~1214 has well-spaced active longitudes, its true rotation period may be a higher 
multiple of 53~days (e.g., $\approx$100~days).

\citet{nfi+2013} monitored GJ~1214 for stellar variability over the relatively short timespan of 78 days  in 2012 with the MITSuMe 0.50~m telescope in Japan and found a shorter period of 44.3 days.  Additional photometric monitoring in 2012 and 2013 was reported by \citet{nms+2015} who used the 1.2~m twin robotic telescopes STELLA (STELLar  Activity) located on Tenerife in the Canary Islands.  They found possible periods of 83.0, 69.0, and 79.6 days.  \citet{mhj+2018} continued long-term monitoring of GJ~1214 with STELLA to create light curves from 2012 through 2016 primarily in the Johnson $BV$ pass bands.  Their most significant signal  was $125\pm5$ days for the 2014-2016 $B+V$ data set, which they claimed overrules previous suggestions of a significantly shorter stellar rotation period.

\begin{figure*}[t]
\begin{centering}
\includegraphics[width=0.6\textwidth, angle =270 ]{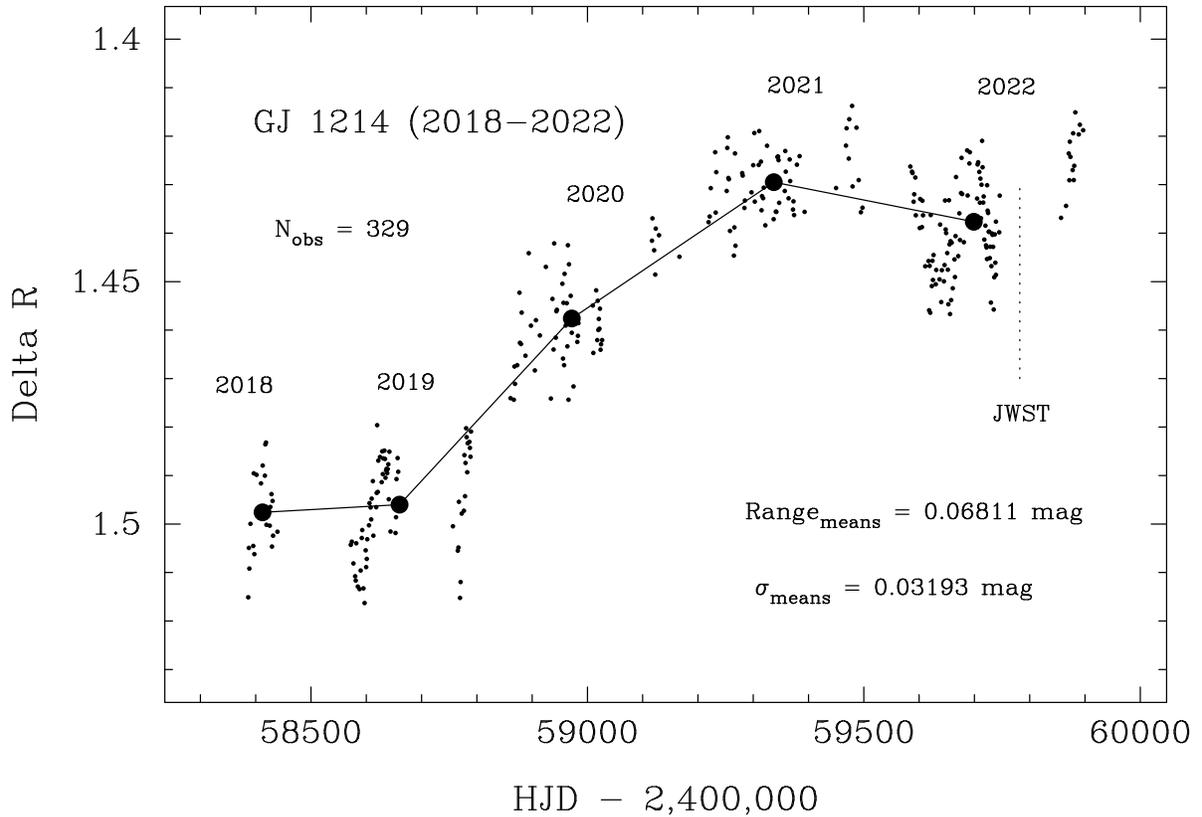}
\caption{Nightly Cousins $R$ band photometry of GJ~1214 from the five 
observing seasons 2018 through 2022 (small circles), acquired with the C14 
automated imaging telescope (AIT) at Fairborn Observatory.  The star is 
slightly variable within each observing season over a range of 1-2\%.  The seasonal
means are plotted as the large filled circles and show that GJ~1214b has 
brightened by several percent over the course of our observations (see 
Table~\ref{table}). The time of the \textit{JWST} phase curve observation of GJ~1214b that was obtained July 20 -- 22, 2022 by \citet{bean21b} is indicated by the dashed line.}
\label{fig:timeseries} 
\end{centering}
\end{figure*}

\section{Observations} \label{sec:obs}
In an attempt to determine the correct rotation period of GJ~1214, we 
conducted our own photometric observations with the Tennessee State 
University Celestron 14-inch (C14) automated imaging telescope (AIT) located 
at Fairborn Observatory in southern Arizona.  We acquired 329 good nightly 
photometric observations (excluding occasional transit observations) during 
the five observing seasons 2018 through 2022.  The observations were 
made through a Cousins $R$ filter with an SBIG STL-1001E CCD camera.  Each 
nightly observation consists of 3--5 consecutive exposures of the GJ~1214 
field of view.  The individual frames are co-added and reduced to differential 
magnitudes in the sense GJ~1214 minus the mean brightness of 13 constant 
comparison stars in the same field.  Further details of our observing and 
data reduction procedures can be found in \citet{sws+2015}.

The nightly observations are plotted as small filled circles in Figure~\ref{fig:timeseries}.
Since GJ~1214 comes to opposition with the Sun on June 11, much of the 
observing season occurs during our annual Summer Shutdown when all telescopes
must be closed from early July to early September due to the ``monsoon season'' 
in southern Arizona.  Thus, there are gaps in each year's light curve when 
no observations can collected.  The 2018 observing season is an exception 
since our initial observations did not take place until after the 2018 
Shutdown.  The yearly mean differential observations are also plotted in 
Figure~\ref{fig:timeseries} as the large filled circles and include the observations on both 
sides of the Summer Shutdown.  The uncertainties in the seasonal means are
roughly the size of the plot symbols.

The observations are summarized by season in Table~\ref{table}.  The standard deviations
of the individual observations from their respective seasonal means are given 
in column~4 and range from 6.44 to 9.96~mmag.  The typical 
precision of
a single nightly observation with the C14 AIT is 2--3~mmag on good 
nights \citep[e.g.,][]{fdm+2021}, so the standard deviations given in column~4
indicate the presence of low-level, night-to-night brightness variability in 
GJ~1214 during each observing season.  The seasonal means given in column~5 
cover a range of 68~mmag, showing a general brightening trend in GJ~1214 of 
several percent over our five years of observation.  Finally, we performed 
period analyses of the individual yearly light curves using the procedure 
described in \citet{wck+2022}.  The resulting best periods are given in 
column~6 and range between 56.4 and 99.6 days. The period listed for 2018 is 
particularly uncertain due to the low number of observations.  Phase curves 
of the five observing seasons are plotted with their five individual periods 
in Figure~\ref{fig:folded}.  Peak-to-peak amplitudes are given in each panel and range 
from 10 to 20~mmag.  Like previously published results, we are unable to identify confidently the true stellar rotation period of 
GJ~1214.

\section{Conclusion} \label{sec:con}
\begin{figure*}
\begin{centering}
\epsscale{1.0}
\plotone{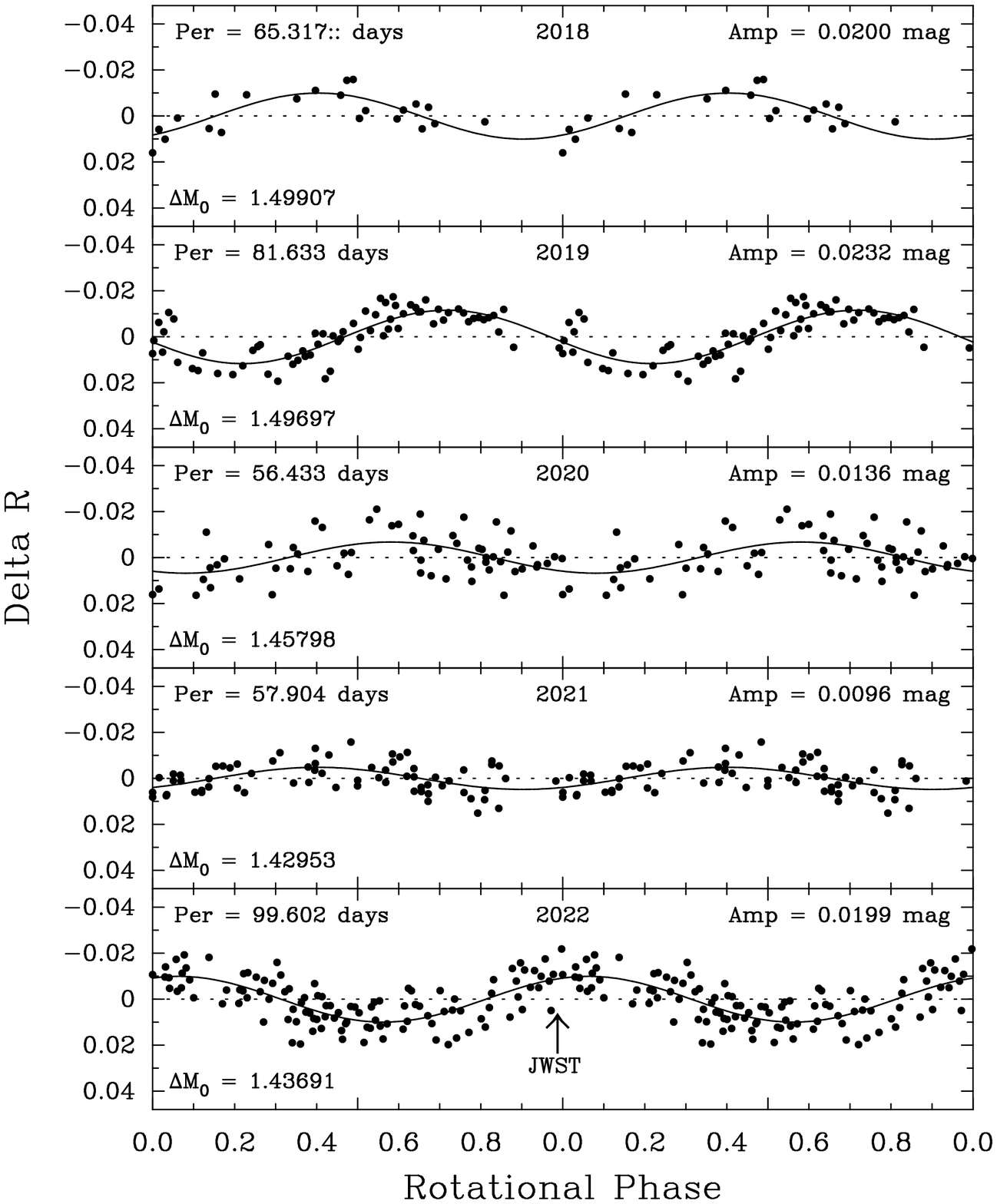}
\caption{Cousins $R$ band phase curves of GJ~1214 from 2018 to 2022.  Periods 
for each year are given in the upper left of each panel.  Peak-to-peak 
amplitudes are given in the upper right.  Like the results given in the 
literature, our periods cover a broad range.  The arrow in the bottom
panel marks the phase of the July 20-22, 2022 \textit{JWST} observation.}
\label{fig:folded} 
\end{centering}
\end{figure*}

We can, however, use our photometric results to predict the starspot coverage at the time of the \textit{JWST} MIRI/LRS phase curve observation of GJ~1214b that was taken in July 2022 \citep{bean21b}.  \citet{rlh+2018} examined the patterns of brightness variation for 72 Sun-like stars and demonstrated that the brightness variability in young, active (and therefore convective) stars is driven by dark spots in the sense that a star is fainter when it is more active (spotted).  The combination of Figure~\ref{fig:timeseries} and the bottom panel of Figure~\ref{fig:folded} show that GJ~1214 was near a long-term as well as a short-term brightness maximum.  In other words, the star was near starspot minimum at the time of the \textit{JWST} phase curve observation. Continued photometric monitoring would be valuable to provide context for the upcoming NIRCam transmission spectroscopy observations of \citet{greene17}, which are currently scheduled for July and August 2023

\begin{deluxetable*}{cccccc}
\label{table} 
\tablewidth{0pt}
\tablecaption{SUMMARY OF C14 AIT PHOTOMETRY OF GJ 1214}
\tablehead{
\colhead{Observing} & \colhead{} & \colhead{Date Range} & \colhead{Sigma} &
\colhead{Seasonal Mean} & \colhead{Period} \\
\colhead{Season} & \colhead{$N_{obs}$} & \colhead{(HJD $-$ 2,400,000)} &
\colhead{(mag)} & \colhead{(mag)} & \colhead{(days)} \\
\colhead{(1)} & \colhead{(2)} & \colhead{(3)} & \colhead{(4)} &
\colhead{(5)} & \colhead{(6)}
}
\startdata
   2018   &  22 & 58386--58439 & 0.00815 & 1.49759 & $\approx65::$ \\
   2019   &  70 & 58571--58789 & 0.00983 & 1.49599 & $81.6\pm1.6$ \\
   2020   &  61 & 58861--59166 & 0.00934 & 1.45758 & $56.4\pm0.6$ \\
   2021   &  67 & 59219--59497 & 0.00644 & 1.42948 & $57.9\pm0.6$ \\
   2022   & 109 & 59584--59896 & 0.00996 & 1.43765 & $99.6\pm1.7$ \\
\enddata
\end{deluxetable*}



\begin{thebibliography}{}
\expandafter\ifx\csname natexlab\endcsname\relax\def\natexlab#1{#1}\fi
\providecommand{\url}[1]{\href{#1}{#1}}
\providecommand{\dodoi}[1]{doi:~\href{http://doi.org/#1}{\nolinkurl{#1}}}
\providecommand{\doeprint}[1]{\href{http://ascl.net/#1}{\nolinkurl{http://ascl.net/#1}}}
\providecommand{\doarXiv}[1]{\href{https://arxiv.org/abs/#1}{\nolinkurl{https://arxiv.org/abs/#1}}}

\bibitem[{{Bean} {et~al.}(2010){Bean}, {Miller-Ricci Kempton}, \&
  {Homeier}}]{bean10}
{Bean}, J.~L., {Miller-Ricci Kempton}, E., \& {Homeier}, D. 2010, \nat, 468,
  669, \dodoi{10.1038/nature09596}

\bibitem[{{Bean} {et~al.}(2021{\natexlab{a}}){Bean}, {Raymond}, \&
  {Owen}}]{bean21a}
{Bean}, J.~L., {Raymond}, S.~N., \& {Owen}, J.~E. 2021{\natexlab{a}}, Journal
  of Geophysical Research (Planets), 126, e06639, \dodoi{10.1029/2020JE006639}

\bibitem[{{Bean} {et~al.}(2011){Bean}, {D{\'e}sert}, {Kabath}, {Stalder},
  {Seager}, {Miller-Ricci Kempton}, {Berta}, {Homeier}, {Walsh}, \&
  {Seifahrt}}]{bean11}
{Bean}, J.~L., {D{\'e}sert}, J.-M., {Kabath}, P., {et~al.} 2011, \apj, 743, 92,
  \dodoi{10.1088/0004-637X/743/1/92}

\bibitem[{{Bean} {et~al.}(2021{\natexlab{b}}){Bean}, {Kempton}, {Fu}, {Gao},
  {Ih}, {Kataria}, {Malik}, {Mansfield}, {Parmentier}, {Rauscher}, {Roman},
  {Stevenson}, \& {Taylor}}]{bean21b}
{Bean}, J.~L., {Kempton}, E. M.~R., {Fu}, G., {et~al.} 2021{\natexlab{b}},
  {Unlocking the Mysteries of the Archetype Sub-Neptune GJ1214b with a
  Full-Orbit Phase Curve}, JWST Proposal. Cycle 1, ID. \#1803

\bibitem[{{Berta} {et~al.}(2011){Berta}, {Charbonneau}, {Bean}, {Irwin},
  {Burke}, {D{\'e}sert}, {Nutzman}, \& {Falco}}]{bcb+2011}
{Berta}, Z.~K., {Charbonneau}, D., {Bean}, J., {et~al.} 2011, \apj, 736, 12,
  \dodoi{10.1088/0004-637X/736/1/12}

\bibitem[{{Berta} {et~al.}(2012){Berta}, {Charbonneau}, {D{\'e}sert},
  {Miller-Ricci Kempton}, {McCullough}, {Burke}, {Fortney}, {Irwin}, {Nutzman},
  \& {Homeier}}]{berta12}
{Berta}, Z.~K., {Charbonneau}, D., {D{\'e}sert}, J.-M., {et~al.} 2012, \apj,
  747, 35, \dodoi{10.1088/0004-637X/747/1/35}

\bibitem[{{Carter} {et~al.}(2011){Carter}, {Winn}, {Holman}, {Fabrycky},
  {Berta}, {Burke}, \& {Nutzman}}]{cwh+2011}
{Carter}, J.~A., {Winn}, J.~N., {Holman}, M.~J., {et~al.} 2011, \apj, 730, 82,
  \dodoi{10.1088/0004-637X/730/2/82}

\bibitem[{{Charbonneau} {et~al.}(2009){Charbonneau}, {Berta}, {Irwin}, {Burke},
  {Nutzman}, {Buchhave}, {Lovis}, {Bonfils}, {Latham}, {Udry}, {Murray-Clay},
  {Holman}, {Falco}, {Winn}, {Queloz}, {Pepe}, {Mayor}, {Delfosse}, \&
  {Forveille}}]{cbi+2009}
{Charbonneau}, D., {Berta}, Z.~K., {Irwin}, J., {et~al.} 2009, \nat, 462, 891,
  \dodoi{10.1038/nature08679}

\bibitem[{{Croll} {et~al.}(2011){Croll}, {Albert}, {Jayawardhana},
  {Miller-Ricci Kempton}, {Fortney}, {Murray}, \& {Neilson}}]{croll11}
{Croll}, B., {Albert}, L., {Jayawardhana}, R., {et~al.} 2011, \apj, 736, 78,
  \dodoi{10.1088/0004-637X/736/2/78}

\bibitem[{{D{\'e}sert} {et~al.}(2011){D{\'e}sert}, {Bean}, {Miller-Ricci
  Kempton}, {Berta}, {Charbonneau}, {Irwin}, {Fortney}, {Burke}, \&
  {Nutzman}}]{desert11}
{D{\'e}sert}, J.-M., {Bean}, J., {Miller-Ricci Kempton}, E., {et~al.} 2011,
  \apjl, 731, L40, \dodoi{10.1088/2041-8205/731/2/L40}

\bibitem[{{Fraine} {et~al.}(2013){Fraine}, {Deming}, {Gillon}, {Jehin},
  {Demory}, {Benneke}, {Seager}, {Lewis}, {Knutson}, \&
  {D{\'e}sert}}]{fraine13}
{Fraine}, J.~D., {Deming}, D., {Gillon}, M., {et~al.} 2013, \apj, 765, 127,
  \dodoi{10.1088/0004-637X/765/2/127}

\bibitem[{{Fu} {et~al.}(2021){Fu}, {Deming}, {May}, {Stevenson}, {Sing},
  {Lothringer}, {Wakeford}, {Nikolov}, {Mikal-Evans}, {Bourrier}, {dos Santos},
  {Alam}, {Henry}, {Mu{\~n}oz}, \& {L{\'o}pez-Morales}}]{fdm+2021}
{Fu}, G., {Deming}, D., {May}, E., {et~al.} 2021, \aj, 162, 271,
  \dodoi{10.3847/1538-3881/ac3008}

\bibitem[{{Fulton} {et~al.}(2017){Fulton}, {Petigura}, {Howard}, {Isaacson},
  {Marcy}, {Cargile}, {Hebb}, {Weiss}, {Johnson}, {Morton}, {Sinukoff},
  {Crossfield}, \& {Hirsch}}]{fulton17}
{Fulton}, B.~J., {Petigura}, E.~A., {Howard}, A.~W., {et~al.} 2017, \aj, 154,
  109, \dodoi{10.3847/1538-3881/aa80eb}

\bibitem[{{Greene} {et~al.}(2017){Greene}, {Beatty}, {Rieke}, \&
  {Schlawin}}]{greene17}
{Greene}, T.~P., {Beatty}, T.~G., {Rieke}, M.~J., \& {Schlawin}, E. 2017,
  {Transit Spectroscopy of Mature Planets}, JWST Proposal. Cycle 1, ID. \#1185

\bibitem[{{Kasper} {et~al.}(2020){Kasper}, {Bean}, {Oklop{\v{c}}i{\'c}},
  {Malsky}, {Kempton}, {D{\'e}sert}, {Rogers}, \& {Mansfield}}]{kasper20}
{Kasper}, D., {Bean}, J.~L., {Oklop{\v{c}}i{\'c}}, A., {et~al.} 2020, \aj, 160,
  258, \dodoi{10.3847/1538-3881/abbee6}

\bibitem[{{Kreidberg} {et~al.}(2014){Kreidberg}, {Bean}, {D{\'e}sert},
  {Benneke}, {Deming}, {Stevenson}, {Seager}, {Berta-Thompson}, {Seifahrt}, \&
  {Homeier}}]{kreidberg14}
{Kreidberg}, L., {Bean}, J.~L., {D{\'e}sert}, J.-M., {et~al.} 2014, \nat, 505,
  69, \dodoi{10.1038/nature12888}

\bibitem[{{Mallonn} {et~al.}(2018){Mallonn}, {Herrero}, {Juvan}, {von Essen},
  {Rosich}, {Ribas}, {Granzer}, {Alexoudi}, \& {Strassmeier}}]{mhj+2018}
{Mallonn}, M., {Herrero}, E., {Juvan}, I.~G., {et~al.} 2018, \aap, 614, A35,
  \dodoi{10.1051/0004-6361/201732300}

\bibitem[{{Narita} {et~al.}(2013){Narita}, {Fukui}, {Ikoma}, {Hori},
  {Kurosaki}, {Kawashima}, {Nagayama}, {Onitsuka}, {Sukom}, {Nakajima},
  {Tamura}, {Kuroda}, {Yanagisawa}, {Hirano}, {Kawauchi}, {Kuzuhara}, {Ohnuki},
  {Suenaga}, {Takahashi}, {Izumiura}, {Kawai}, \& {Yoshida}}]{nfi+2013}
{Narita}, N., {Fukui}, A., {Ikoma}, M., {et~al.} 2013, \apj, 773, 144,
  \dodoi{10.1088/0004-637X/773/2/144}

\bibitem[{{Nascimbeni} {et~al.}(2015){Nascimbeni}, {Mallonn}, {Scandariato},
  {Pagano}, {Piotto}, {Micela}, {Messina}, {Leto}, {Strassmeier}, {Bisogni}, \&
  {Speziali}}]{nms+2015}
{Nascimbeni}, V., {Mallonn}, M., {Scandariato}, G., {et~al.} 2015, \aap, 579,
  A113, \dodoi{10.1051/0004-6361/201425350}

\bibitem[{{Orell-Miquel} {et~al.}(2022){Orell-Miquel}, {Murgas}, {Pall{\'e}},
  {Lamp{\'o}n}, {L{\'o}pez-Puertas}, {Sanz-Forcada}, {Nagel}, {Kaminski},
  {Casasayas-Barris}, {Nortmann}, {Luque}, {Molaverdikhani}, {Sedaghati},
  {Caballero}, {Amado}, {Bergond}, {Czesla}, {Hatzes}, {Henning},
  {Khalafinejad}, {Montes}, {Morello}, {Quirrenbach}, {Reiners}, {Ribas},
  {S{\'a}nchez-L{\'o}pez}, {Schweitzer}, {Stangret}, {Yan}, \& {Zapatero
  Osorio}}]{orell22}
{Orell-Miquel}, J., {Murgas}, F., {Pall{\'e}}, E., {et~al.} 2022, \aap, 659,
  A55, \dodoi{10.1051/0004-6361/202142455}

\bibitem[{{Radick} {et~al.}(2018){Radick}, {Lockwood}, {Henry}, {Hall}, \&
  {Pevtsov}}]{rlh+2018}
{Radick}, R.~R., {Lockwood}, G.~W., {Henry}, G.~W., {Hall}, J.~C., \&
  {Pevtsov}, A.~A. 2018, \apj, 855, 75, \dodoi{10.3847/1538-4357/aaaae3}

\bibitem[{{Rogers} \& {Seager}(2010)}]{rogers10}
{Rogers}, L.~A., \& {Seager}, S. 2010, \apj, 716, 1208,
  \dodoi{10.1088/0004-637X/716/2/1208}

\bibitem[{{Sing} {et~al.}(2015){Sing}, {Wakeford}, {Showman}, {Nikolov},
  {Fortney}, {Burrows}, {Ballester}, {Deming}, {Aigrain}, {D{\'e}sert},
  {Gibson}, {Henry}, {Knutson}, {Lecavelier des Etangs}, {Pont},
  {Vidal-Madjar}, {Williamson}, \& {Wilson}}]{sws+2015}
{Sing}, D.~K., {Wakeford}, H.~R., {Showman}, A.~P., {et~al.} 2015, \mnras, 446,
  2428, \dodoi{10.1093/mnras/stu2279}

\bibitem[{{Spake} {et~al.}(2022){Spake}, {Oklop{\v{c}}i{\'c}}, {Hillenbrand},
  {Knutson}, {Kasper}, {Dai}, {Orell-Miquel}, {Vissapragada}, {Zhang}, \&
  {Bean}}]{spake22}
{Spake}, J.~J., {Oklop{\v{c}}i{\'c}}, A., {Hillenbrand}, L.~A., {et~al.} 2022,
  \apjl, 939, L11, \dodoi{10.3847/2041-8213/ac88c9}

\bibitem[{{Van Eylen} {et~al.}(2018){Van Eylen}, {Agentoft}, {Lundkvist},
  {Kjeldsen}, {Owen}, {Fulton}, {Petigura}, \& {Snellen}}]{vaneylen18}
{Van Eylen}, V., {Agentoft}, C., {Lundkvist}, M.~S., {et~al.} 2018, \mnras,
  479, 4786, \dodoi{10.1093/mnras/sty1783}

\bibitem[{{Wong} {et~al.}(2022){Wong}, {Chachan}, {Knutson}, {Henry}, {Adams},
  {Kataria}, {Benneke}, {Gao}, {Deming}, {L{\'o}pez-Morales}, {Sing}, {Alam},
  {Ballester}, {Barstow}, {Buchhave}, {dos Santos}, {Fu}, {Garc{\'\i}a
  Mu{\~n}oz}, {MacDonald}, {Mikal-Evans}, {Sanz-Forcada}, \&
  {Wakeford}}]{wck+2022}
{Wong}, I., {Chachan}, Y., {Knutson}, H.~A., {et~al.} 2022, \aj, 164, 30,
  \dodoi{10.3847/1538-3881/ac7234}

\end{thebibliography}
\end{document}